\documentclass[twocolumn,showpacs,aps, prb, floatfix, amsmath]{revtex4}
\usepackage{dcolumn}
\usepackage{bm}
\usepackage{graphicx}
\usepackage{bm}
\usepackage{color}

\begin{document}
\textbf{Comment on \textquotedblleft{ Reversibility of magnetic field driven transition from electronic phase separation state to single-phase state in manganites: A microscopic view} \textquotedblright}

\vspace{5mm}

The paper being commented on \cite{Liu2017} reports Magnetic Force Microscopy (MFM) measurement on (La$_{2/3}$Pr$_{1/3}$)$_{5/8}$Ca$_{3/8}$MnO$_3$ (LPCMO) thin film and claim to address the reversibility of charge order insulator (COI)- ferromagnetic metal (FMM) transition first time. Contrary to this claim, there are many reports on the irreversibility of COI-FMM transition in LPCMO bulk (polycrystal, single crystal) as well as thin film.  It includes systematic investigation by our group on the path dependence of COI-FMM transition in LPCMO film using macroscopic measurement as well as MFM measurement as a function of magnetic field and temperature.\cite{Mishra2015, Rawat2013, Mishra2013, Sathe2010} None of these references has been cited by the authors. 

Our work \cite{Mishra2015, Rawat2013, Mishra2013, Sathe2010} has shown that COI-FMM transition in case of LPCMO film (on NGO, STO and LAO substrate) is interrupted and high temperature phase remains arrested on cooling in the absence of externally applied magnetic field. This arrested phase can be transformed to FMM phase by applying magnetic field isothermally. With reducing magnetic field to zero the field induced FMM phase is retained and results in open loop in isothermal magnetoresistance (MR) at low temperature. We have also investigated the metastability of low temperature state using CHUF (cooling and heating in unequal magnetic field) protocol \cite{Banerjee2009}, which clearly brings out the metastable nature of COI phase at low temperature. Before our work in LPCMO thin film, the studies on bulk LPCMO system also clearly demarcated the region of thermomagnetic irreversibility and its nature \cite{Kumar2006, Wu2006, Ghivelder2005} (reference 8 and 9 cited as reference 11, 25 of the paper being commented on). These result contradicts the authors claim that \textbf{\textquotedblleft While It has been well known that a magnetic field can drive the transition of the EPS state into a single-phase state in manganites, the reversibility of this transition is not well studied.\textquotedblright} (see abstract \cite{Liu2017}).

The claim that previous MFM studies only followed the initial curve is also not true e.g. see page 1 second column paragraph two \cite{Liu2017} where author state, \textbf{\textquotedblleft We note that previous MFM studies only follow the initial magnetization process [11,12,21]. The reversed process, i.e., whether the system would recover the initial state after the field is reduced, remains unknown.\textquotedblright} Contrary to this statement, reference \cite{Zhou2015} (or reference [12] in the above quote) shows MFM measurement corresponding to complete MH loop and initial magnetization curve (e.g. see figure 2 of Zhou et al. \cite{Zhou2015}) At the same time our MFM work on LPCMO thin film as a function of temperature as well as magnetic field has been ignored \cite{Rawat2013}.  Our work \cite{Rawat2013} showed that field induced FMM state is retained at 100 K on reducing magnetic field to zero. It clearly states that images during second field increasing cycle are similar to that observed during field reducing cycle (page 3, last paragraph of Rawat et al. \cite{Rawat2013})

Ultimately author interpret their result in terms of glass freezing e.g. \textbf{\textquotedblleft The latter thought argues that glass freezing blocks the COI-to-FMM transition in the middle stage leading to the EPS state at low temperatures.\textquotedblright} (page 6, first column, second para) and at the end of their paper, \textbf{\textquotedblleft Our observation is also highly valuable for the application of manganite materials in spintronics as it gives clear guidance on how to achieve EPS state or single FMM state by design \textquotedblright} The idea of glass freezing has been proposed and tested in a wide variety of system ( see Chaddah \cite{Chaddah2015} and references therein),and it includes LPCMO in bulk \cite{Kumar2006, Wu2006} as well as thin films \cite{Mishra2015, Rawat2013, Mishra2013, Sathe2010}. These works also showed that it is possible to continuously tune the phase fraction of two phases, from zero to hundred percent, at low temperature.

In the abstract also these authors \cite{Liu2017} stated,\textbf{\textquotedblleft We argue that EPS state is a consequence of system quenching whose response to an external magnetic field is governed by a local energy landscape.\textquotedblright}  The use of word system quenching hints that the authors have vague idea of quench disorder. Earlier reports have clearly showed that EPS is a consequence of quenched disorder, which can be imparted by doping as well as strain disorder \cite{Mishra2015}. In fact the EPS (and as consequence the local energy landscape) can be controlled by tuning strain disorder in LPCMO films grown on STO \cite{Mishra2015, Mishra2013}. Further, previous reports showed that the two phases present in the EPS are structurally dissimilar \cite{Mishra2015}, and used the CHUF protocol to establish that phase fractions present at a particular temperature and applied magnetic field could be controlled. Thus previous reports had exhaustively explained many of the claims of the present paper and provided comprehensive understanding of the physics involved in the EPS.

We wish to point out a serious oversight in the paper being commented on \cite{Liu2017}. Both figure 2(c) and figure 3(a) show the initial magnetization and MH loop at 10 K. In case of figure 3(a) initial curve clearly lies outside the envelope curve, whereas in case of figure 2(c) no such distinction is reported.

\vspace{5mm}
      \textbf{R Rawat, V. G. Sathe and P Chaddah}
      
     UGC-DAE Consortium for Scientific Research, University Campus, Khandwa Road, Indore-452001, India.

{}

\end{document}